\documentclass[aps,prl,showpacs,twocolumn]{revtex4}%
\usepackage{amsfonts}
\usepackage{amssymb}
\usepackage{amsmath}
\usepackage{graphicx}
\usepackage{dcolumn}
\usepackage{bm}
\usepackage{units}
\usepackage{color}%
\providecommand{\U}[1]{\protect\rule{.1in}{.1in}}
\begin{document}
\title{Thermal spin-transfer in Fe-MgO-Fe tunnel junctions}
\author{Xingtao Jia}
\affiliation{Department of Physics, Beijing Normal University, Beijing 100875, China}
\author{Ke Xia}
\affiliation{Department of Physics, Beijing Normal University, Beijing 100875, China}
\author{Gerrit E. W. Bauer}
\affiliation{Institute for Materials Research, Tohoku University, Sendai 980-8577, Japan}
\affiliation{Delft University of Technology, Kavli Institute of NanoScience, 2628 CJ Delft,
The Netherlands}
\date{\today }

\begin{abstract}
We compute thermal spin transfer torques (TST) in Fe-MgO-Fe tunnel
junctions using a first principles wave function-matching method. At
room
temperature, the TST in a junction with 3 MgO monolayers amounts to 10$^{-7}%
\operatorname{J}%
$/$%
\operatorname{m}%
^{2}/%
\operatorname{K}%
$, which is estimated to cause magnetization reversal for temperature
differences over the barrier of the order of 10 $%
\operatorname{K}%
$. The large TST  can be explained by multiple scattering between interface
states through ultrathin barriers. The angular dependence of the TST can be
very skewed, possibly leading to thermally induced high-frequency generation.

\end{abstract}

\maketitle

Spin-dependent thermoelectric effects in metallic magnetic systems
have been known for quite some time \cite{johnson1987} but recently
experience renewed interest. Spin caloritronic phenomena
\cite{Bauer2010} include the spin Seebeck effect \cite{SSE}, which
should be distinguished from the spin-dependent Seebeck effect in
nanostructures \cite{Slachter}. Large spin-related Peltier cooling
effects have been measured in magnetic NiCu nanopillars
\cite{Sugihara10}. Hatami \textit{et al}. \cite{hatami2007}
predicted that a temperature gradient induces a spin transfer torque
that can excite a magnetization. Experimental evidence for the
thermal spin-transfer torque has been obtained for Co-Cu-Co
nanowires \cite{Haiming10}. Slonczewski recently argued that thermal
torques can be generated efficiently in spin valves with polarizing
magnetic insulators \cite{Slonczewski2010}.

Magnetic tunnel junctions (MTJs) of transition metals with MgO barriers
\cite{Yuasa04,Parkin04} have great potential for applications in magnetic
random access memory (MRAM) elements and high-frequency generators
\cite{Oh2009,Deac2008,Jung2010,Matsumoto2009}. An important goal of academic
and corporate research remains the reduction of the critical currents
necessary to induce magnetization precession and reversal \cite{Yoda10,Ikeda}.
Spin-dependent Seebeck effects in MTJs have very recently been computed
\cite{Heiliger} and measured \cite{Muenzenberg}. A spin accumulations has been
injected thermally into silicon by permalloy contacts through MgO tunnel
junctions \cite{Jansen}.

Here we predict very large thermal spin transfer torques in MTJs
with thin MgO barriers, which might open new possibilities to design
memory elements and high-frequency generators driven by heat
currents only. We have been motivated by the strong energy
dependence of electron transmission through MTJs with thin barriers
due to the existence of interface resonant states
\cite{surface-state}, which should cause large thermoelectric
effects. Focussing on epitaxial Fe-MgO-Fe MTJs under a temperature
bias, we demonstrate the effectiveness of thermal spin transfer
torques by \textit{ab initio} calculations based on the
Landauer-B\"{u}ttiker transport formalism.
\begin{figure}[h]
\includegraphics[width=8.0cm]{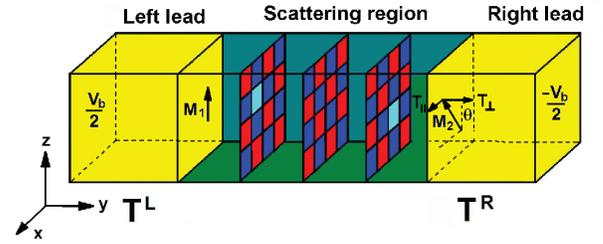}\caption{Schematic Fe-MgO(3ML)-%
Fe(001) MTJs. We consider both a temperature difference $\Delta T$
and voltage difference $V_{b}$\ between the ferromagnetic
reservoirs. The magnetization $\mathbf{M}_{1}$ of the left lead is
fixed along the $z$-axis, while the magnetization $\mathbf{M}_{2}$
of the right lead is rotated by an angle $\theta$ in the $xz$ plane
relative to $\mathbf{M}_{1}$. The small dark gray (red and dark
blue) squares in the scattering region represent O and Mg atoms,
respectively, while the light gray (light blue) ones denote oxygen
vacancies.}
\label{scheme}%
\end{figure}
Consider an MTJ as sketched in Fig. \ref{scheme} with a voltage and
temperature bias over the two leads, which are in local thermal
equilibrium with Fermi-Dirac distribution functions $f_{L/R}\left(
\epsilon\right) =[e^{(\epsilon-\mu_{L/R})/k_{B}T_{L/R}}+1]^{-1}$ and
local chemical potentials $\mu_{L/R}$ and temperatures $T_{L/R}$.
The generalized Landauer-B\"{u}ttiker formalism~\cite{butcher1990}
is very suitable to handle transport through layered magnetic
structures. The spin current from the $n$-th layer to the $n+1$-th
layer can be written as \cite{wang2008}
\begin{equation}
\mathbf{J}_{n+1,n}=\frac{1}{8\pi}\int d\epsilon\left[  \mathbf{t}_{n+1,n}%
^{L}(\epsilon)f_{L}(\epsilon)+\mathbf{t}_{n,n+1}^{R}(\epsilon)f_{R}%
(\epsilon)\right]  .\label{eq:spincurrent}%
\end{equation}
Here the energy-dependent spin transmission coefficient matrix from the left
(right) direction is defined as $\mathbf{t}_{n+1,n}^{L/R}(\epsilon
)=\sum_{\mathbf{k}_{\parallel}}\langle\Psi_{\mathbf{k}_{\parallel}}%
^{L/R}(\epsilon)|\hat{\mathcal{J}}_{n+1,n}(\mathbf{k}_{\parallel}%
)|\Psi_{\mathbf{k}_{\parallel}}^{L/R}(\epsilon)\rangle$ with spin current
operator $\hbar\hat{\mathcal{J}}_{n+1,n}(\mathbf{k}_{\parallel})=-%
\operatorname{Re}%
\sum_{L,L^{\prime}}\left\{  \mathbf{\hat{\sigma},}\hat{H}_{nL,n+1L^{\prime}%
}(\mathbf{k}_{\parallel})\right\}  $, $\hat{H}_{nL,n+1L^{\prime}}%
(\mathbf{k}_{\parallel})$ denotes the Hamiltonian matrix in spin
space~\cite{wang2008}, where $L\equiv(l,m)$ are the azimuthal and magnetic
quantum numbers. $\mathbf{\hat{\sigma}}$ is the vector of Pauli spin matrices
and $\mathbf{k}_{\Vert}$ is integrated over the two-dimensional Brillouin zone
of transverse modes.

When the applied voltage vanishes and $k_{B}\Delta T\ll\epsilon_{F}$ we may
expand Eq. (\ref{eq:spincurrent}) in $\Delta T=T_{R}-T_{L}.$%
\begin{align}
\mathbf{J}_{n+1,n}  & =\frac{1}{8\pi}\{\int d\epsilon f(\epsilon_{F}%
,T_{0})[\mathbf{t}_{n+1,n}^{L}(\epsilon)+\mathbf{t}_{n,n+1}^{R}(\epsilon
)]\nonumber\\
& +\frac{\Delta T}{2T_{0}}\int d\epsilon(\epsilon-\epsilon_{F})\frac{\partial
f}{\partial\epsilon}[\mathbf{t}_{n+1,n}^{L}(\epsilon)-\mathbf{t}_{n,n+1}%
^{R}(\epsilon)]\},\label{TST_full}%
\end{align}
where $T_{0}\equiv(T^{L}+T^{R})/2$ . The first term in $\mathbf{J}_{n+1,n}$
$=$ $\mathbf{J}_{n+1,n}^{eq}+\mathbf{J}_{n+1,n}^{\Delta T}$ is the equilibrium
spin current that communicates the exchange coupling through the barrier,
while $\mathbf{J}_{n+1,n}^{\Delta T}$ is the thermal spin current. The torque
acting on the $n$-th layer is the difference between the incoming and outgoing
spin currents $\mathbf{T}_{n}^{\Delta T}=\mathbf{J}_{n,n-1}^{\Delta T}$ $-$
$\mathbf{J}_{n+1,n}^{\Delta T}$. The total thermal spin transfer (TST) torque
is then
\begin{equation}
\mathbf{T}_{\Delta T}=\frac{\Delta T}{eT_{0}}\int d\epsilon(\epsilon
-\epsilon_{F})\mathbf{\tau}_{V}(\epsilon)\frac{\partial}{\partial\epsilon
}f(\epsilon),\label{thermal_torque}%
\end{equation}
where $\mathbf{\tau}_{V}(\epsilon)=(e/16\pi)\sum_{n=N}^{\infty}[\mathbf{t}%
_{n,n-1}^{L}(\epsilon)-\mathbf{t}_{n-1,n}^{R}(\epsilon)-\mathbf{t}_{n+1,n}%
^{L}(\epsilon)+\mathbf{t}_{n,n+1}^{R}(\epsilon)]$ is the electrical torkance,
and the sum from $N$ to $\infty$\ runs from the first layer at the interface
until deep into the bulk of the \textquotedblleft free\textquotedblright%
\ magnetic lead. The thermal torkance $\mathbf{\tau}_{T}=\mathbf{T}_{\Delta
T}/\Delta T$ depends on the energy dependent transmission and $T_{0}$. Only
when $\mathbf{\tau}_{V}(\epsilon)$ varies slowly around the Fermi level in the
thermal window $k_{B}\Delta T$, the Sommerfeld expansion may be employed and
$\mathbf{\tau}_{T}\rightarrow-(e/2)\left(  \pi^{2}k_{B}^{2}T_{0}/6\right)
\partial\mathbf{\tau}_{V}(\epsilon)|_{\epsilon_{F}}$. For comparison, the
linear-response voltage-driven torkance reads \cite{wang2008} $\mathbf{T}%
_{\Delta V}\Delta V(\epsilon_{F})/V_{b}\rightarrow\mathbf{\tau}_{V}%
(\epsilon_{F}),$ where $V_{b}$ is the applied bias. We found that Eq.
(\ref{thermal_torque}) accurately reproduces non-linear calculations based on
Eq. (\ref{eq:spincurrent}) in the parameter regime considered here.

Here we focus mainly on Fe-MgO-Fe(100) with 3 layers (3$L$) MgO ($\sim6\,%
\operatorname{\text{\AA}}%
$), corresponding to the thinnest barrier that can be reliably grown
\cite{Oh2009,Deac2008,Jung2010,Matsumoto2009}. Ignoring the small lattice
mismatch between the lead and barriers, we assume an isomorphous structure
with unrelaxed interfacial atoms at bcc positions\textit{.}\ We use a
$1200\times1200$ $k$-point mesh in the full two-dimensional Brillouin zone
(BZ) of transport channels to ensure numerical convergence. Details of the
electronic structure and transport calculations can be found elsewhere
\cite{ke10}.

\begin{figure}[ptb]
\includegraphics[width=8.0cm]{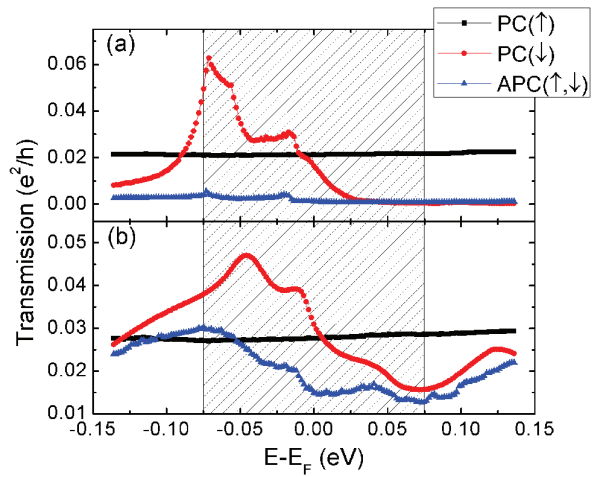}\caption{Energy-dependent transmission
of Fe-MgO(3$L$)-Fe(001) MTJs with (a) perfect interfaces and (b)
10\% OVs at both interfaces. The shaded area indicates the thermal
energy window
$k_{B}T_{0}$ at room temperature.}%
\label{3lmgo_trans}%
\end{figure}

The energy-dependent transmission through our MTJ is plotted in Fig.
\ref{3lmgo_trans} for the parallel (P) and antiparallel (AP) configurations.
Vacancies break the crystalline symmetry \cite{miao2008}, broaden the resonant
peaks of the minority spin channel, and enhance the coupling between majority
$\Delta_{1}$ state of one lead and the surface states of the minority spin on
the other side. The newly opened channels increase the AP conductance.

For specular Fe-MgO interfaces, we find a zero-bias
\textquotedblleft optimistic\textquotedblright\ tunneling
magnetoresistance (TMR) of $1300\%$
and a resistance area $0.063\,%
\operatorname{\Omega }%
\operatorname{\mu m}%
^{2}$. Our TMR ratio is consistent with a \textquotedblleft
pessimistic\textquotedblright\ TMR calculated to be around 0.93 for the same
barrier thickness \cite{Mathon2001} with majority-spin transmission around
$0.4e^{2}/h$ \cite{Wortmann2004}. When 10\% oxygen vacancies (OVs) (the
energetically most favorable defects) are introduced at the interface, the TMR
decreases to 96\% and $RA=0.036\,%
\operatorname{\Omega }%
\operatorname{\mu m}%
^{2}$. When 10\% OVs are introduced in the middle of barrier, the TMR drops to
70\%, and the $RA$ slightly increases to $0.076\,%
\operatorname{\Omega }%
\operatorname{\mu m}%
^{2}$. These results are comparable with the measured $0.19\,%
\operatorname{\Omega }%
\operatorname{\mu m}%
^{2}$. and $\mathrm{TMR}=15\%$ for a similar barrier thickness at room
temperature \cite{Matsumoto2009}.

From the energy-dependent transmissions, we can also compute the magneto
Seebeck coefficients \cite{Heiliger} and electronic heat conductances. At
$T=10%
\operatorname{K}%
$, we find $\varkappa_{e}^{\mathrm{P}}=2.1\times10^{6}%
\operatorname{W}%
\operatorname{K}%
^{-1}%
\operatorname{m}%
^{-2}$ and $\varkappa_{e}^{\mathrm{AP}}=0.14\times10^{6}%
\operatorname{W}%
\operatorname{K}%
^{-1}%
\operatorname{m}%
^{-2}$, which, including estimated phonon contributions to the heat
conductance, leads to the thermoelectric figure of merit $\left(  ZT\right)
_{10%
\operatorname{K}%
}\simeq10^{-3}.$

\begin{figure}[tb]
\includegraphics[width=8.5cm]{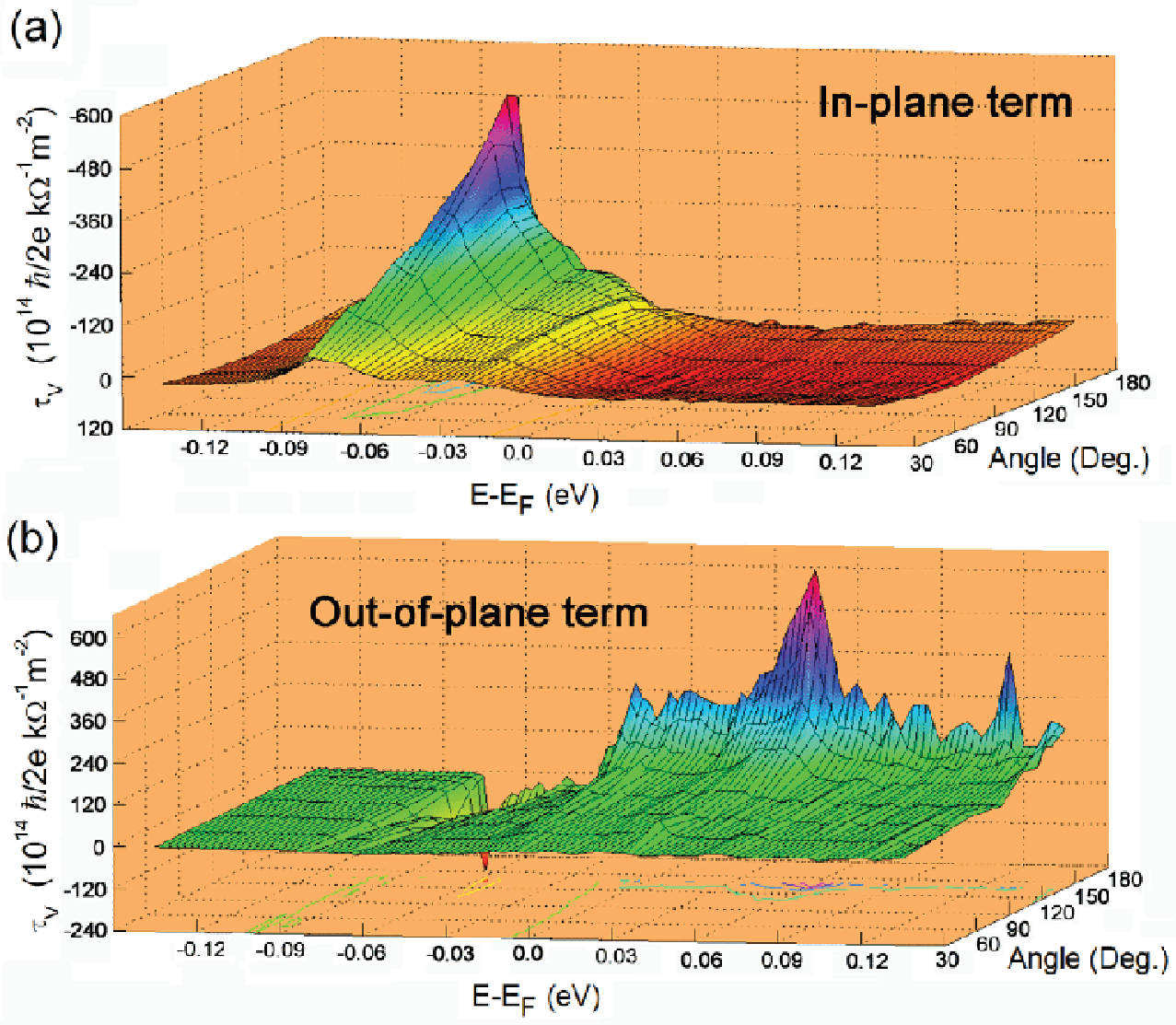}
\includegraphics[width=8.0cm]{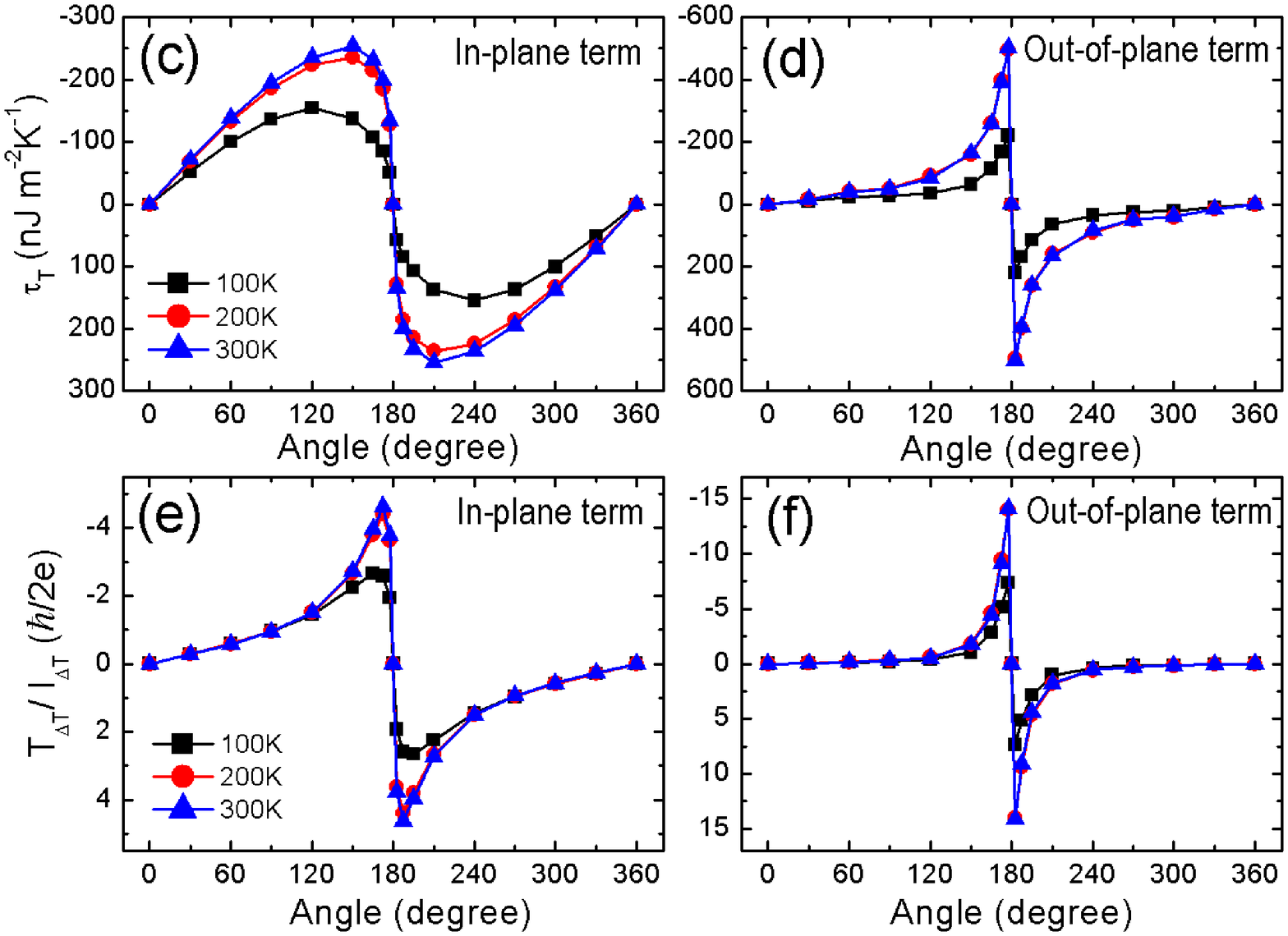}\caption{Energy and angle-dependent
(a) in-plane and (b) out-of-plane torkance, angular dependent (c) in-plane and
(d) out-of-planeTST, and ratio of (e) in-plane and (f) out-of-plane TST to
thermocurrent ($T_{\Delta T}$$/$$I_{\Delta T}$) of epitaxial Fe-%
MgO(3$L$)-Fe(001) MTJs.}%
\label{3lmgo_torque}%
\end{figure}

In Fig. \ref{3lmgo_torque} (a) and (b), we present the in- and
out-of-plane angular-resolved torkances of specular Fe-MgO(3$L$)-Fe
MTJs. The in-plane torkance is smooth in most energy regions,
indicating good numerical
convergence. We observe two resonances: a small one at $E_{F}-0.02\,%
\operatorname{eV}%
$ that contributes to the TST for $T_{0}\gtrsim100\,%
\operatorname{K}%
$ and a sharp and larger peak at $E_{F}-0.0725\,%
\operatorname{eV}%
$ that contributes to the TST for $T_{0}\gtrsim300\,%
\operatorname{K}%
$. The out-of-plane torkance is much more sensitive to numerical details. The
noise in Fig. \ref{3lmgo_torque} does not affect the integrated TST, however.
At energies far away from the Fermi level ($E\geq E_{F}+0.03\,%
\operatorname{eV}%
$ and $E\leq E_{F}-0.09\,%
\operatorname{eV}%
$), the in-plane torkance is small and proportional to $\sin\theta$
as predicted by model studies \cite{Butler2006}. However, this
region contributes only weakly to the TST. The two sharp peaks near
the Fermi level show an angular dependence that deviates strongly
from a sine function. The asymmetry of the angular dependence of the
in-plane TST reflects multiple scattering in the barrier and is
therefore exponentially suppressed for thick layers. The in-plane
TST of 7$L$ MTJ (not shown) already agrees well with a sine
function.

The angular dependence of the observable TST, \textit{i.e.} the energy
integral in Eq. (\ref{thermal_torque}), is plotted in Figs. \ref{3lmgo_torque}%
(c) and (d). We observe strong deviations from a sine function at all
temperatures considered. The skewness can be traced to multiple-reflection
hot-spots caused by the interfacial resonances mentioned above. At room
temperature the in-plane torkance peaks around 165$%
\operatorname{{{}^\circ}}%
$ and the functional form can be fitted to an asymmetry parameter
\cite{Slonczewski2002} of $\Lambda=3.5$. This value is much larger than
observed for the voltage-induced torque in metallic spin valves \cite{Rippard}%
, which should be beneficial for high-frequency generation \cite{Rippard}. We
therefore suggest the possibility of efficient spin oscillators driven by heat
flows through MTJs. The out-of-plane term is an effective field that dominates
the in-plane term for angles $>$ 165$%
\operatorname{{{}^\circ}}%
$.

Fig. \ref{3lmgo_torque}(e) and (f) displays the angular dependence
of the spin transfer efficiency monitored by the ratio of the
torque to charge current
density for a given temperature bias $\Delta T=1%
\operatorname{K}%
$. We find that the ratio (for both in-plane and out-of-plane terms)
increases strongly close to the APC, for which the charge current is
suppressed by the high spin polarization of the Fe-MgO interface
\cite{Slonczewski1996}.

\begin{figure}[ptb]
\includegraphics[width=8.5cm]{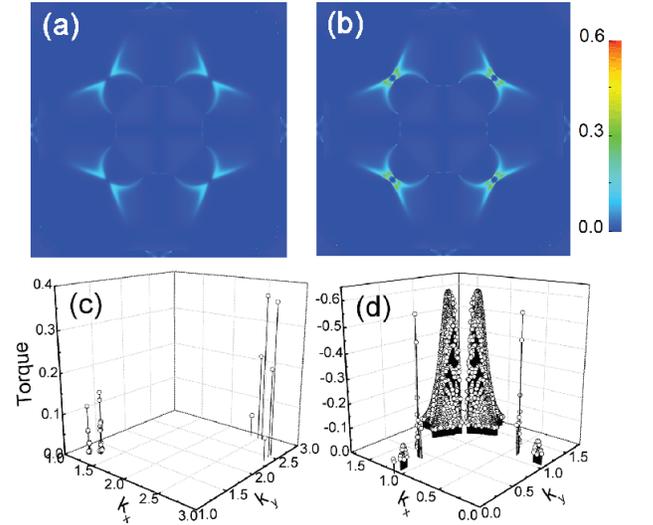}
\caption{$\mathbf{k}_{\parallel}$-resolved electron transmission
probability of (a) majority- and (b) minority-spin, and in-plane
torkance from (c) majority- and (d) minority-spin channels in 3L
MTJs with magnetization angle of 177.5$\operatorname{{{}^\circ}}$ at
energy $E_{F}-0.0725\operatorname{eV}$. Majority (minority) are
defined for the left lead. The integrated transmission probabilities
are $4.6 (6.3)\times 10^{-3}$ $e^{2}/h$ and integrated in-plane
torkances are $5.5\times10^{-5}$
($-4.2\times10^{-3})\operatorname{eV}$ for
majority (minority) spins, respectively.}%
\label{fig4}%
\end{figure}

The high spin transfer efficiency near the APC can be explained by multiple
reflection due to resonant tunneling. In Fig. \ref{fig4} resonant tunneling is
observed in the APC at a chosen energy in both spin channels with a
conductance polarization of 16\%.
Their contributions to the torkance are much larger, since the minority spin
channel transfers 99\% of the torkance due to its high interfacial electronic
density of states. Here majority and minority spins are defined for the left
lead. The resonance persists in the exact APC (Fig. \ref{3lmgo_trans}(a)) but
spin transfer vanishes for collinear magnetizations.

In Table \ref{tab_mgo_thermal} we compare TSTs equivalent to $\Delta T=1\,%
\operatorname{K}%
$ at $T_{0}=300\,%
\operatorname{K}%
$ with electric STs for MTJs with 3, 5, and 7 MgO layers. The equivalent bias
and current density of the thinnest barrier sample is much larger than that of
the thicker one, which reflects the exponential decay of the conductance as a
function of barrier thickness. $\Delta V_{eq}$ is the ratio of thermal to
electric torkance, which is larger for 3-MgO ($1\,%
\operatorname{K}%
\sim0.27\,\mathrm{m}%
\operatorname{eV}%
$). $\Delta V_{1K}$ demonstrates that TSTs decrease faster than the electric
STs when the barrier gets thicker. The sign change in $\Delta V_{1\mathrm{K}}$
as a function of barrier thickness is attributed to that of the Seebeck
coefficient. Moreover, the torque to current density ratio $T/I$ is larger for
the thermal than the electric case, indicating the superior efficiency of spin
angular moment transfer by temperature differences.

\begin{table}[ptb]
\caption{Thermal torque $T_{\mathrm{1K}}$ per unit cell in $nL$ MgO
MTJs at $T=300\,\operatorname{K}$ and $\Delta T=1\,\operatorname{K}$
under closed and
open (in brackets) circuit conditions for $\theta=90\operatorname{{{}^\circ}}%
$. $\Delta V_{\mathrm{eq}}=$ $T_{\mathrm{1K}}/\tau_{_{V}}$ is the equivalent
bias. $\Delta V_{1\mathrm{K}}$/$I_{1\mathrm{K}}$ is the
thermovoltage/thermocurrent, $T_{V}$ and $I_{V}$ are electrically induced
torque and current, respectively.}%
\begin{tabular*}
{8.5cm}[c]{@{\extracolsep{\fill}}lcccccc}\hline\hline
$n$ & $\tau_{_{V}}$ & $T_{1K}$* & $\Delta V_{eq}$ & $\Delta
V_{1\operatorname{K}}$ & $T_{1K}/I_{1K}$ & $T_{V}/I_{V}$\\
& ($\operatorname{mJ}/\operatorname{V}/\operatorname{m}^{2}$) &
($\operatorname{nJ}\operatorname{m}^{-2}$) & ($\operatorname{mV}$) &
($\operatorname{mV}$) & ($\hbar/2e$) & ($\hbar/2e$)\\\hline
3 & 0.72 & -195(-232) & -0.27 & 0.052 & -0.94 & 0.21\\
5 & 0.082 & -3.32(-5.33) & -0.040 & 0.025 & -0.84 & 0.46\\
7 & 0.011 & -0.24(-0.062) & -0.021 & -0.0154 & 3.58 & 0.98\\\hline\hline
\multicolumn{7}{l}{{*}$1\,\operatorname{J}\operatorname{V}^{-1}%
\operatorname{m}^{-2}=3\times 10^{18}(\hbar/2)\operatorname{k}\Omega
^{-1}\operatorname{m}^{-2}$}%
\end{tabular*}
\label{tab_mgo_thermal}%
\end{table}

The TST is potentially useful for manipulating the magnetic configurations in
MTJs with thin barriers. We estimate the critical temperature bias $\Delta
T_{c}$ by comparing the TST with the measured torques at the critical voltage
biases in CoFeB MTJ's at room temperature \cite{Wang2011}. For 3ML MgO the
switch from APC to PC should occur close to $\Delta T_{c}^{\mathrm{AP}%
\rightarrow\mathrm{P}}=6.5\,%
\operatorname{K}%
$
since then $\left\vert \mathbf{T}_{\Delta T}/\sin\theta\right\vert
=20\times10^{-6}\,%
\operatorname{J}%
\operatorname{m}%
^{-2}$ equals the critical torque for electric switching. At $\Delta
T_{c}^{\mathrm{P}\rightarrow\mathrm{AP}}$ $=56.5\,%
\operatorname{K}%
$, $\left\vert \mathbf{T}_{\Delta T}/\sin\theta\right\vert =8.2\times10^{-6}\,%
\operatorname{J}%
\operatorname{m}%
^{-2} $ equals the critical torque for electric PC to APC switching
\cite{Wang2011}. We note that $\tau_{T}$ is function of the global temperature
that saturates around $275%
\operatorname{K}%
$. Room temperature conditions are therefore favorable for thermal
magnetization switching.

In an open circuit, the thermoelectric current vanishes, but not the
thermospin current, thereby allowing transfer of angular momentum without
transfer of charge. The thermal torque is even found to be larger in the
closed compared to the open circuit, since the equivalent bias $\Delta
V_{\mathrm{eq}}$ and the thermovoltage $\Delta V_{\mathrm{1K}}$ have opposite
signs.

\begin{figure}[h]
\includegraphics[width=8.5cm]{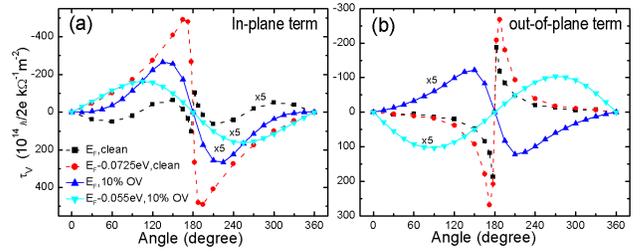}\caption{Angular dependent (a) in-plane
and (b) out-of-plane torkance in Fe-MgO(3$L$)-Fe(001) MTJs with
specular and disordered interfaces at two selected energies, {
i.e.}, at the Fermi energy and at the resonance. Squares and circles
are results for specular interfaces at $E_{F}$ and
$E_{F}-0.0725\,\operatorname{eV}$, respectively; the up-pointing
triangles and down-pointing triangles are disordered samples at
$E_{F}$ and $E_{F}-0.055\,\operatorname{eV}$, respectively.}
\label{disorder}%
\end{figure}

The spectral features due to resonances are sensitive to disorder.
In Fig. \ref{disorder} we show the angular-dependent torkance in
3-MgO with 10\% OVs at both interfaces. We make comparison for two
situations: one is at $E_{F}$, and another is at resonant peaks near
to $E_{F}$. The resonant peaks in the clean samples at
$E_{F}-0.0725\,\operatorname{eV}$ shift to lower energy (around
$E_{F}-0.055\,\operatorname{eV}$) in the presence of 10\% OV as
shown in Fig. \ref{3lmgo_trans}, so different energies are chosen to
compare clean and dirty situations. We observe that the disorder to
a large extent restores the $\sin\theta$ angular dependence. The
order of magnitude of the in-plane torkance of the ideal junctions
at $E_{F}$ is unmodified. The situation at the resonant peak is more
complicated by a shift from $E_{F}-0.0725\,\operatorname{eV}$
(specular) to $E_{F}-0.055\,\operatorname{eV}$ (disordered) with
decreased amplitude. A full calculation of the TST in the presence
of OV disorder at room temperature is beyond our present
computational capacity, but the two noted changes of the resonance
will at least partly cancel each other.

In summary, we calculate TSTs of the order of
$10^{-6}\,\operatorname{J}\operatorname{m}^{-2} $ in ultrathin
Fe-MgO-Fe tunnel junctions at room temperature for a temperature
bias of $10\,\operatorname{K}$. A strong asymmetric angular
dependence of TSTs is predicted for ballistic junctions. Based on
these results we predict heat-flow-induced magnetization reversal
and high frequency generation in magnetic tunnel junctions.

We gratefully acknowledge financial support from National Basic Research
Program of China (973 Program) under the grant No. 2011CB921803, NSF-China
grant No. 60825404, the Dutch FOM Foundation and EU-ICT-7 contract no. 257159 MACALO.

\end{document}